\documentclass{IEEEtran}
\usepackage{cite}
\usepackage{amsmath,amssymb,amsfonts}
\usepackage{graphicx}
\usepackage{textcomp,nicefrac}
\usepackage{comment}
\usepackage[hidelinks]{hyperref}
\usepackage[OT1]{fontenc} 
\usepackage{tikz}
\newcommand\copyrighttext{%
  \footnotesize \textcopyright \the\year{} IEEE. Personal use of this material is permitted. Permission from IEEE must be obtained for all other uses, including reprinting/republishing this material for advertising or promotional purposes, collecting new collected works for resale or redistribution to servers or lists, or reuse of any copyrighted component of this work in other works.}

\newcommand\copyrightnotice{%
\begin{tikzpicture}[remember picture,overlay]
\node[anchor=south,yshift=10pt] at (current page.south) {\fbox{\parbox{\dimexpr0.75\textwidth-\fboxsep-\fboxrule\relax}{\copyrighttext}}};
\end{tikzpicture}%
}

\renewcommand\fbox{\fcolorbox{red}{white}}
\def\BibTeX{{\rm B\kern-.05em{\sc i\kern-.025em b}\kern-.08em
T\kern-.1667em\lower.7ex\hbox{E}\kern-.125emX}}
\begin{document}
\title{A High Compression Ratio Channel Multiplexing Method for Micro-pattern Gaseous Detectors}
\author{Yu Wang, Shubin Liu*, Hao Zhuang, Zhengwu Ding, Zhihang Yao, Changqing Feng, Zhiyong Zhang
\thanks{Manuscript received 20 May 2024; revised 26 Dec 2024. This work was supported by the Young Scientists Fund of the National Natural Science Foundation (Grant No. 12205297) 
and the National Science Fund for Distinguished Young Scholars(Grant No. 12025504)}
\thanks{Yu Wang, Shubin Liu, Hao Zhuang, Zhengwu Ding, Zhihang Yao, Changqing Feng and Zhiyong Zhang are with State Key Laboratory of Particle Detection and Electronics University of Science and Technology of China, No.96, Jinzhai Road, Hefei, Anhui, China (email: liushb@ustc.edu.cn).
Yu Wang, Shubin Liu, Hao Zhuang, Zhengwu Ding, Changqing Feng and Zhiyong Zhang are also with the Department of Modern Physics, University of Science and Technology of China, Hefei 230026, China.\par
Zhihang Yao and Shubin Liu are also with School of Nuclear Science and Technology, University of Science and Technology of China, Hefei 230026, China.
}
}

\maketitle
\copyrightnotice

\begin{abstract}
The demand for a large number of readout channels has been a limiting factor for the application of Micro-pattern Gaseous Detectors (MPGDs) in achieving higher spatial resolution and larger detection areas. This challenge is further compounded by issues related to system integration, power consumption, and cost efficiency.
To address these challenges, this study proposes two novel multiplexing methods based on Eulerian circuits. 
Mathematical calculations indicate that with $n$ electronics channels, up to $n \times (n-1)/2 - (n - 2)/2 + 1$ detector channels can be read out, where $n$ is even.
Three types of multiplexing circuits were designed, implemented, and tested in combination with Micromegas detectors. 
Experimental results demonstrate that, for a multiplexing circuit with a factor of 8, the spatial resolution remains comparable to the direct readout method, while achieving a detection efficiency exceeding 94\%. 
For a circuit with a multiplexing factor of 16, although the spatial resolution shows a slight degradation, the detection efficiency remains above 93.6\%.
These results demonstrate that the proposed multiplexing methods can significantly reduce the number of readout channels while maintaining an acceptable level of spatial resolution and detection efficiency.
These findings highlight the potential of the proposed multiplexing techniques for applications in fields requiring high-resolution and cost-effective detector systems, such as cosmic-ray muon imaging. 
\end{abstract}

\begin{IEEEkeywords}
Detector readout, Micro-pattern gas detectors, Multi-channel high-density signal readout and integration, Multiplexing readout
\end{IEEEkeywords}

\section{INTRODUCTION}
\label{sec:introduction}
\IEEEPARstart{M}{icro}-pattern gaseous detectors (MPGDs) are a class of gaseous ionization detectors characterized by sub-millimeter avalanche structures. 
Prominent examples include gas electron multiplier (GEM)~\cite{bib_gem}, micro-mesh gaseous structures (Micromegas) detector~\cite{bib_mm}, thick GEM (THGEM)~\cite{bib_thgem}, and micro-resistance well ($\mu$RWELL) detector~\cite{bib_rwell}. 
MPGDs are widely recognized for their ability to achieve sub-millimeter spatial resolution while being scalable to large detection areas, often exceeding sub-square meter dimensions.

To achieve high spatial resolution, the geometric dimensions of the readout units in MPGDs must be minimized.
For strip readout, the strip width typically needs to be no more than 4 to 6 times the target spatial resolution.
For instance, achieving a spatial resolution better than $100~\mathrm{\mu m}$ typically requires a readout strip width of approximately $400~\mathrm{\mu m}$. 
In this case, an MPGD with an area of $100~\mathrm{cm} \times 100~\mathrm{cm}$ would require 2500 readout channels for single-dimensional strip readout.
However, the demand for such a large number of readout channels significantly increases system complexity, posing challenges in integration, power consumption, cooling, and cost, particularly for large-scale applications.

To address these challenges, three main approaches have been proposed.
Firstly, the development of highly integrated and compact front-end electronics, together with corresponding data acquisition systems, enables efficient signal readout in a compact design.
Secondly, the design of the readout unit can be improved by modifying the structure of the readout strips into a zigzag shape~\cite{Perez_Lara_a_comparative_study_of_straight_strip_and_zigzag_interleaved}. 
This approach takes advantage of the fact that ionization charges diffuse while drifting to the anode. 
By applying the charge-centroid method, the distance between strips can be increased. 
Thirdly, the channel multiplexing method connects multiple detector readout strips to a single readout electronics channel, significantly reducing the required number of readout channels~\cite{bib_procureur_genatic_multiplexing,bib_Yue_an_encoding_readout_method,bib_Yue_Mathematical_modelling_and_study,bib_BQi_A_novel_method_of_encoded_multiplexing_readout_2016,bib_GYuan_2D_encoded_multiplexing_readout_for_thgem}.

Channel multiplexing has demonstrated significant potential in addressing the challenges posed by the large number of readout channels.
In experiments with low channel occupancy rates, where a typical event activates only a few adjacent detector channels, most electronics channels remain idle.
Channel multiplexing exploits this sparsity by reusing a limited number of electronics channels across multiple readout strips, thereby substantially reducing the total number of required channels. 
The core principle of channel multiplexing is to compress the number of readout channels by using the detector's known signal distribution and spatial characteristics of the events.
Common implementations of this approach include inductive-encoded multiplexing readout and position-encoded multiplexing readout.

Several multiplexing methods have been developed in previous studies.
Procureur at Saclay~\cite{bib_procureur_genatic_multiplexing} designed a circuit that multiplexed 1024 strips into 61 electronics channels. However, this study lacked a comprehensive method for constructing the multiplexing scheme.
Yue at Tsinghua University~\cite{bib_Yue_an_encoding_readout_method,bib_Yue_Mathematical_modelling_and_study} and our group~\cite{bib_BQi_A_novel_method_of_encoded_multiplexing_readout_2016,bib_GYuan_2D_encoded_multiplexing_readout_for_thgem,bib_JLiu_An_encoding_readout_scheme_for_micromegas_detector_used_in_muography,bib_JPan_position_encoding_readout_electronics_of_large_area2019} proposed basic models for channel multiplexing and validated them on small-area detectors.
These studies have demonstrated the effectiveness of channel multiplexing in significantly reducing the number of required readout electronics channels.
However, further research is required to introduce additional constraints to the basic multiplexing model to ensure the unique reconstruction of hit positions and the uniformity of reuse factors, particularly when applied to large-area MPGDs.

To address this limitation, this study proposes two novel multiplexing methods, supported by corresponding mathematical models.
Based on these methods, we designed and implemented various multiplexing circuits and validated their performance using self-designed thermal-bonding Micromegas detectors.
By employing a channel multiplexing method with a compression ratio of 16:1, a detector with 1024 strips can be read out using only 64-channel front-end electronics, with the noise level increasing from $0.45~f\mathrm{C}$ to $0.81~f\mathrm{C}$. 
Although the gain of the front-end electronics is affected by the larger input capacitance, the detector efficiency decreases slightly from 96.7\% to 93.5\% compared with direct readout. 
Moreover, we integrated these multiplexing circuits with compact readout electronics to construct large-scale muon imaging facilities, demonstrating the effectiveness of the multiplexing circuits in experiments with low channel occupancy rates.

\section{Principle of the Encoded Multiplexing Method}
\subsection{Basic Model of Encoded Multiplexing}
The position encoded multiplexing method relies on two fundamental conditions:
\begin{itemize}
	\item The charge signal from a single event spreads across multiple detector channels
	\item The activated (fired) detector channels are contiguous.
\end{itemize}

These two conditions are easily satisfied because the readout units of MPGDs typically have dimensions on the order of several hundred micrometers. 
Additionally, the diffusion range of primary ionization during drifting is larger than the size of these readout units~\cite{bib_G_aielli_The_rpc_space_resolution_2014,bib_MIodice_Performance_studies_ofMicromegas_2014,bib_LLavezzi_Standalone_codes_for_simulation_and_reconstruction_2020}.

The process of position encoded multiplexing involves pairing consecutive detector channels  {$strip_i,$ $strip_{i+1}$} with a pair of electronics channels {$ch_a$, $ch_b$}. 
Since the detector channels are contiguous, this process can be simplified to mapping a pair of contiguous number ($i$, $i+1$) to a unique pair of number ($a$, $b$). 
This mapping can be formalized as a mathematical model using the concept of Eulerian graphs and Eulerian circuits.  
This type of graph is referred to as an encoded multiplexing graph.

An Eulerian circuit is defined as a path in a graph that traverses each edge exactly once and returns to its starting point.
Under this theoretical framework, electronics channels are modeled as vertices, and detector channels as edges.
This ensures that each pair of detector channels uniquely maps to a pair of electronics channels, avoiding edge repetition.

\begin{figure}[htbp]
\centerline{\includegraphics[width=3.5in]{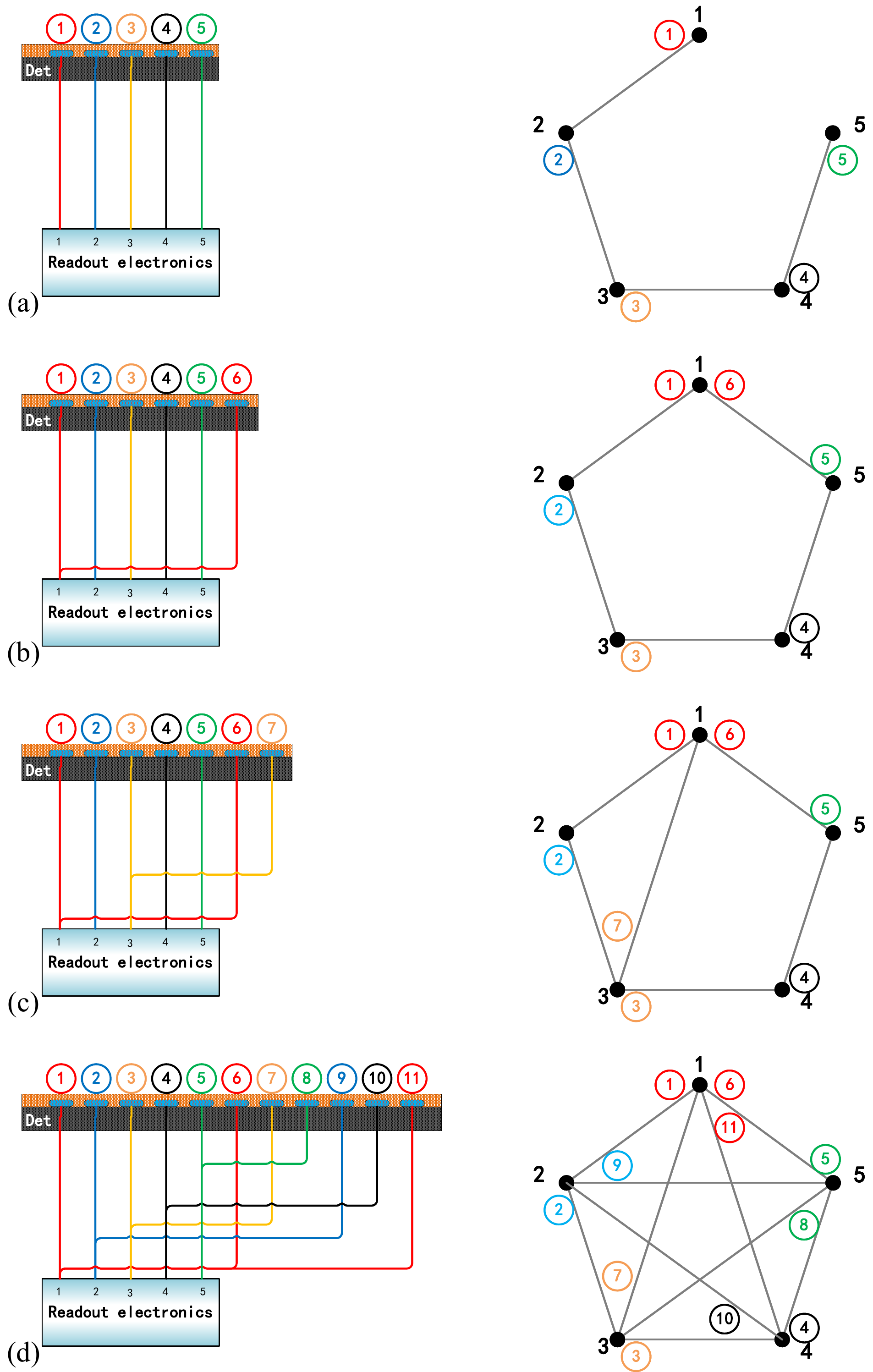}}
\caption{Process of constructing the encoded multiplexing graph. (a) Direct readout scheme.
(b) Scheme for reading six detector channels with five electronics channels, with channel 1 multiplexed.
(c) Scheme for reading seven detector channels with five electronics channels, with channels 1 and 3 multiplexed.
(d) Example of multiplexing five electronics channels to read out eleven detector strips, with all channels multiplexed. }
\label{fig1}
\end{figure}

\figurename~\ref{fig1} illustrates the step-by-step procedure for constructing the encoded multiplexing graph.
In \figurename~\ref{fig1}(a), the left subfigure presents a schematic diagram of a direct readout scheme, where five detector channels are read using five electronics channels. 
The right subfigure shows the corresponding ``encoded multiplexing graph'' for this direct readout. 
The vertices (black dots) represent the electronics channels, with the numbers near them indicating the electronics channel numbers. 
The number inside the circles represent the number of detector channel connected to the respective electronics channel. 
The edges between vertices illustrate the mapping relationships between the electronics and detector channels.

As shown in \figurename~\ref{fig1}(b), when an additional detector channel needs to be read out, it can be connected to electronics channel 1, thereby multiplexing it.
By applying the two conditions mentioned earlier, ambiguity is eliminated because multiple detector channels fire simultaneously, and these channels are contiguous.
For instance, if electronics channels 1 and 5 read out signals, the fired detector channels are 5 and 6.
The right subfigure in \figurename~\ref{fig1}(b) illustrates the encoded multiplexing graph, where contiguous detector channels 5 and 6 are uniquely mapped to electronics channels 5 and 1.

To read out another channel, it can be connected to electronics channel 3, as shown in \figurename~\ref{fig1}(c).
In this case, the contiguous detector channels 6 and 7 are mapped to a unique pair of electronics channels 1 and 3.

By applying the same logic, a schematic diagram illustrating the connection of five electronics channels to read out eleven detector channels is shown on the left side of \figurename~\ref{fig1}(d).
The corresponding Eulerian graph is displayed on the right side of \figurename~\ref{fig1}(d). 
When any two electronics channels are activated, it is possible to infer the unique consecutively hit detector channels.
For example, if electronics channels 2 and 4 readout signals above the threshold, the possible hit positions on the detector are 2, 4, 9, and 10.
Among these, only strips 9 and 10 are consecutive, making them the valid pairs.

Therefore, the encoded multiplexing method can be described using the Eulerian path theorem.
The construction of the encoded scheme involves designing an Eulerian walk that traverses each edge exactly once, where the edges correspond to the contiguous channels of the detector, and the vertices represent the electronics channels.

According to the Eulerian path theorem, a connected graph has an Eulerian cycle if and only if every vertex has an even degree~\cite{bib_DRBean_Recursive_Euler_and_hamilton_paths_1976}.
To maximize the multiplexing factor, the encoded multiplexing graph must connect as many edges as possible between every vertex, with the maximum configuration being a complete graph.
When the number of electronics channels, $n$, is odd, all vertices in the complete graph have an even degree. The total number of edges in this case is $n \times (n-1)/2$, allowing $n$ electronics channels to read out $n \times (n-1)/2$ detector channels.
For an even number of electronics channels, $n$, the maximum number of detector channels that can be read out is calculated as $n \times (n-1)/2 - (n - 2)/2 + 1$.

\subsection{Constraints of the Encoded Multiplexing for Practical Applications}
According to Euler's path theorem, the Eulerian circuit of a specific Eulerian graph is not unique. 
Consequently, the encoded multiplexing scheme corresponding to a given set of readout electronics channels is also not unique. 
The construction of Eulerian circuits is influenced by two factors: the choice of the starting vertex and the order in which the vertices are traversed along the circuit. 
As a result, various multiplexing schemes can be derived depending on the specific Eulerian circuits selected during the design process.
Of these two factors, the choice of the starting vertex does not affect the multiplexing schemes, as all schemes remain equivalent in this regard.
However, the order of vertex traversal significantly impacts the distance between the numbers of two detector channels connected to the same electronics channel. 
This  distance, termed the reuse distance, determines the maximum number of  fired channels that can be uniquely identified without ambiguity.

For example, consider the multiplexing scheme illustrated in \figurename~\ref{fig1}(d), where the reuse distance is 3. In this scheme, electronics channel 5 is multiplexed to detector channels 5 and 8.
If detector channels 5, 6, and 7 are fired, the corresponding electronics channels 1, 3, and 5 will record the signals. 
However, the reconstruction result could correspond to either detector channels 5, 6, and 7 or channels 5, 6, 7, and 8, as there is no additional information to confirm whether detector channel 8 is also fired.

To guarantee the unique reconstruction of hit positions, several constraints must be imposed during the construction of the multiplexing scheme.

Firstly, the minimum distance between the numbers of two detector channels connected to the same electronics channel must be greater than the maximum number of fired channels in a single event.
For high spatial resolution detectors utilizing the channel multiplexing method, it is crucial to maximize the reuse distance, which refers to the interval between repeated connections of electronics channels to detector channels. 
Accurate decoding is only possible if the reuse distance exceeds the maximum number of fired strips in a single event.

Secondly, the multiplexing factor of each electronics channel should be optimized to minimize the capacitance variations induced by the multiplexing of detector channels, as the input capacitance value significantly impacts the gain and noise performance of the front-end electronics.

\subsection{Mathematical Model of Encoded Multiplexing}
To satisfy these constrains, a more practical mathematical model is introduced into the encoded multiplexing method.
Maximizing the reuse distance corresponds to maximizing the number of edges in an Eulerian circuit, a problem known in graph theory as the Eulerian Recurrent Length (ERL) problem.
In graph theory, a circuit is a path that begins and ends at the same vertex~\cite{bib_Reinhard_Diestel_Graph_Theory_2017}, and the length of a circuit is defined as the number of edges in the circuit.
The ERL of an Eulerian graph $G$, denoted as ERL($G$), is the maximum length of the shortest subcycle in any Eulerian circuit of $G$~\cite{bib_Jimbo_Shuji_On_the_Eulerian_recurrent_lengths_of_complete_bipartite_graphs_and_complete_graphs_2014}.

In an encoded multiplexing graph $G$, the ERL($G$) corresponds to the reuse distance, which represents the minimum spacing between repeated uses of a vertex in the graph.
Therefore, the problem of constructing an optimal encoded multiplexing graph is converted to maximizing the ERL of the graph.
This problem has been mathematically addressed by constructing specific Hamiltonian paths for two types of graphs: complete graphs with an odd number of vertices and complete bipartite graphs. For both types of graphs, the ERL is $n - 4$, where $n$ is the number of vertices~\cite{bib_Jimbo_Shuji_The_eulerian_recurrent_length_of_a_complete_graphs_2012, bib_Jimbo_Shuji_On_the_Eulerian_recurrent_lengths_of_complete_bipartite_graphs_and_complete_graphs_2014}. 
Thus, with $n$ electronics channels, the minimum reuse distance is $n - 4$, ensuring efficient resource allocation in the encoded multiplexing graph."

As the number of readout electronics channels is typically even, we have extended the solution to the ERL problem to include complete graphs with an even number of vertices.
The construction equation is presented as:
\begin{equation*}
C_{n}:=\left\{\begin{array}{l}
H_{0} \rightarrow H_{2} \rightarrow \cdots \rightarrow H_{(n-6) / 2} \rightarrow H_{(n-4) / 2} \rightarrow \\
H_{(n-4) / 2-2} \rightarrow \cdots \rightarrow H_{1} \rightarrow n-2, \text { if } n \bmod 4=2 \\
H_{0} \rightarrow H_{2} \rightarrow \cdots \rightarrow H_{(n-4) / 2} \rightarrow H_{(n-6) / 2} \rightarrow \\
H_{(n-6) / 2-2} \rightarrow \cdots \rightarrow H_{1} \rightarrow n-2, \text { if } n \bmod 4=0
\end{array}\right.
\end{equation*}
where  $H_{k}=n-2 \rightarrow v_{0}(k) \rightarrow v_{1}(k) \rightarrow \cdots v_{(n-4) / 2}(k) \rightarrow n-1 \rightarrow v_{(n-2) / 2}(k) \rightarrow \cdots \rightarrow v_{n-3}(k)$ , and $v_i (k)\quad (i \leq n-3)$ is defined as follows:
\begin{equation*}
v_{i}(k)=\left\{\begin{array}{l}
k, \quad i=0 \\
\left(v_{i-1}(k)+i\right) \bmod (n-2), i>0 \text { and } i \bmod 2=1 \\
\left(v_{i-1}(k)-i\right) \bmod (n-2), \text { otherwise }
\end{array}\right.
\end{equation*}

For instance, if ten electronics channels are utilized, the connection relationships can be described as follows. 
In this configuration, the electronics channels are numbered from 0 to 9, and for each arrow, the detector number increases by one.
The relationships are represented as:
\begin{equation*}
 \begin{array}{l}
  8 \rightarrow 0 \rightarrow 1 \rightarrow 7 \rightarrow 2 \rightarrow 9 \rightarrow 6 \rightarrow 3 \rightarrow 5 \rightarrow 4 \rightarrow \\
  8 \rightarrow 2 \rightarrow 3 \rightarrow 1 \rightarrow 4 \rightarrow 9 \rightarrow 0 \rightarrow 5 \rightarrow 7 \rightarrow 6 \rightarrow \\ 
  8 \rightarrow 3 \rightarrow 4 \rightarrow 2 \rightarrow 5 \rightarrow 9 \rightarrow 1 \rightarrow 6 \rightarrow 0 \rightarrow 7 \rightarrow \\ 
  8 \rightarrow 1 \rightarrow 2 \rightarrow 0 \rightarrow 3 \rightarrow 9 \rightarrow 7 \rightarrow 4 \rightarrow 6 \rightarrow 5 \rightarrow 8
  \end{array} 
\end{equation*}

\begin{figure}[htbp]
\centerline{\includegraphics[width=3.5in]{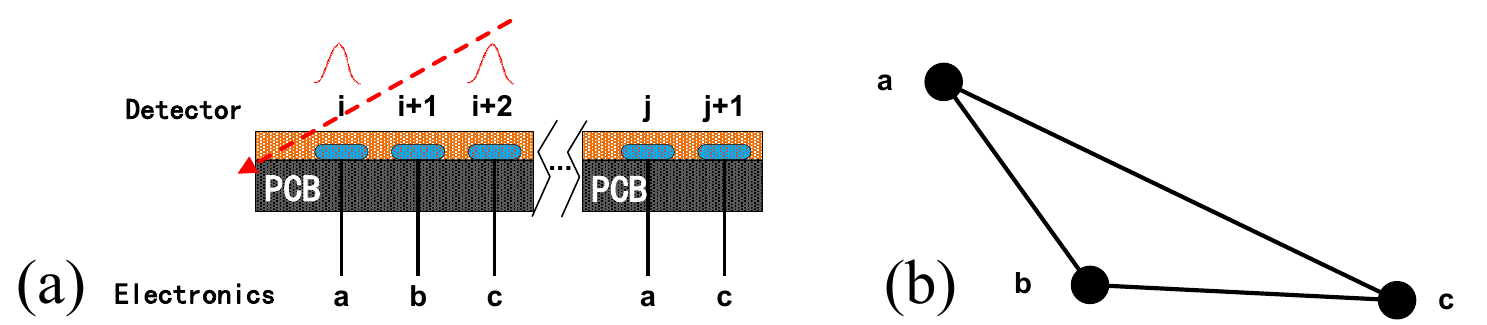}}
\caption{Explanation of the error in position decoding.}
\label{fig2}
\end{figure}

For the bipartite graph, it addresses the issue that arises when certain detector channels fail to record signals.
Consider the scenario illustrated in \figurename~\ref{fig2}(a): in some regions of the encoded multiplexing scheme, electronics channels $a$, $b$, and $c$ are configured to read three contiguous detector channels $i$, $i+1$, and $i+2$. While in another region, electronics channels $a$ and $c$ are designed to read detector channels $j$ and $j+1$.
If a particle hits the detector at channels $i$, $i+1$, and $i+2$, the noise or diffusion of the primary ionization may cause the signal at channel $i+1$ to fall below the detection threshold. Consequently, the read-back data on the electronics side will only include channels $a$ and $c$. 
As a result, the decoded result will correspond to channels $j$ and $j+1$, leading to an error in position measurement.
The cause of this error is illustrated in \figurename~\ref{fig2}(b), where the encoded multiplexing graph contains a triangle, indicating that certain channel combinations are unsuitable for the encoded multiplexing scheme.

\begin{figure}[htbp]
\centerline{\includegraphics[width=3.0 in]{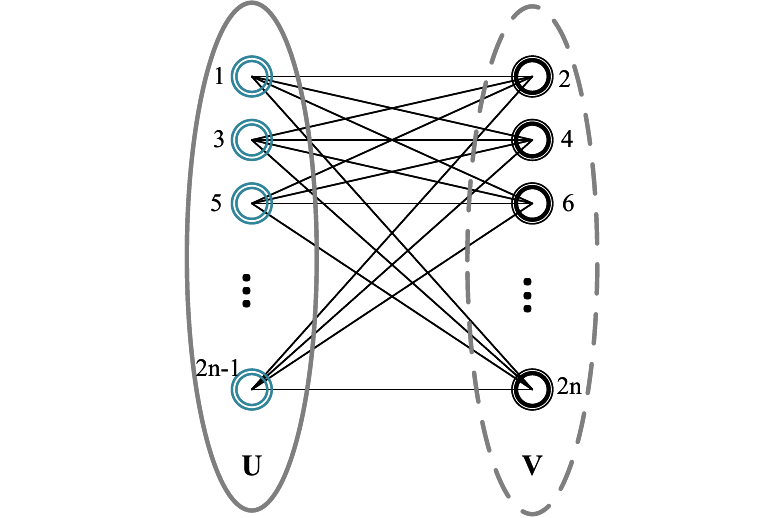}}
\caption{Example of a bipartite graph with $2n$ vertices.}
\label{fig3}
\end{figure}

The bipartite graph provides a solution that eliminates all triangular combinations. 
A bipartite graph is one where vertices are divided into two disjoint sets, with edges only connecting vertices from different sets~\cite{bib_LMetcalf_Chapter5_graph_theory_2016}.
As illustrated in \figurename~\ref{fig3}, the vertices are divided into two groups $U$ and $V$, where no edges exist between vertices within the same group.
Thus, when some channels do not record the signals, the decoded result is none, which would exclude some error results.  
The construction of the Eulerian circuit for bipartite graphs with an even number of vertices is detailed in~\cite{bib_Jimbo_Shuji_On_the_Eulerian_recurrent_lengths_of_complete_bipartite_graphs_and_complete_graphs_2014}.
In such graphs, the Eulerian recurrent length (ERL) is given by $(n-4)$, where $n$ is the number of vertices.

The construction equation of the bipartite graph is shown in (\ref{eq_bipartite_graph}), where $H_{k}$ is defined as in (\ref{eq_halmiton_path_bipartite_graph}). 
In (\ref{eq_halmiton_path_bipartite_graph}), the electronics channels are labeled as $0, 1, 2, \ldots, 2n-1$.
For each transition from one electronics channel to the next, the detector channel index increases by one.
\begin{equation}
 T=H_{0} \rightarrow H_{2} \rightarrow H_{4} \rightarrow \cdots \rightarrow H_{n-2} \rightarrow 0 
  \label{eq_bipartite_graph}
\end{equation}

\begin{equation}
 \begin{aligned} 
 H_{k}= & 0 \rightarrow(2 k+1) \bmod 2 n \rightarrow 2 \rightarrow(2 k+3) \bmod 2 n \\ & \rightarrow \cdots \rightarrow 2 n-2  \rightarrow(2 k+2 n-1) \bmod 2 n
 \end{aligned} 
 \label{eq_halmiton_path_bipartite_graph}
\end{equation}

A C++-based program has been developed to generate the encoded multiplexing scheme for both types of graphs, and the authors have made it publicly available on GitHub~\cite{bib_MyGithub}.

\section{Verification and Test}
\subsection{The Micromegas Detector and the Encoded Multiplexing Circuits}
These two encoded multiplexing methods were tested on a large thermal bonding Micromegas detector with a sensitive area of $400~\mathrm{mm} \times 400~\mathrm{mm}$.
The readout units of the detector are strips. Two orthogonal readout strips are embedded in the second and third layers of the anode readout printed circuit board (PCB). The pitch of the strips is $400~\mu\mathrm{m}$.
The width of the strips in the upper layer is $68~\mu\mathrm{m}$, while that in the lower layer is $295~\mu\mathrm{m}$. 
The drift gap between the drift cathode and the mesh was $5 ~ \mathrm{mm}$, and the avalanche gap was about $100 ~ \mu\mathrm{m}$. 
Further details about the Micromegas detector can be found in~\cite{bib_Jianxingfeng_A_themal_bonding_method_for_manufacturing_Micromegas_detector_2021}, where the detector size was enlarged for this work.

\begin{figure}[htbp]
\centerline{\includegraphics[width=3.5in]{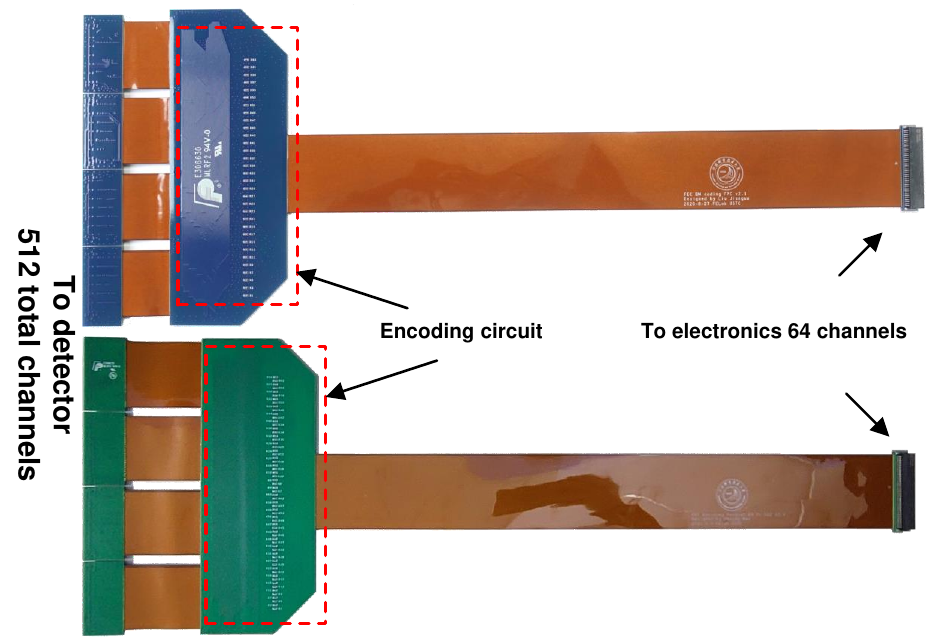}}
\caption{Upper: Circuit multiplexing 512 detector strips to 64 readout electronics using the complete graph method; Lower: Circuit multiplexing 512 detector strips to 64 readout electronics using the bipartite graph method. }
\label{fig4}
\end{figure}

\begin{figure}[htbp]
\centerline{\includegraphics[width=3.5in]{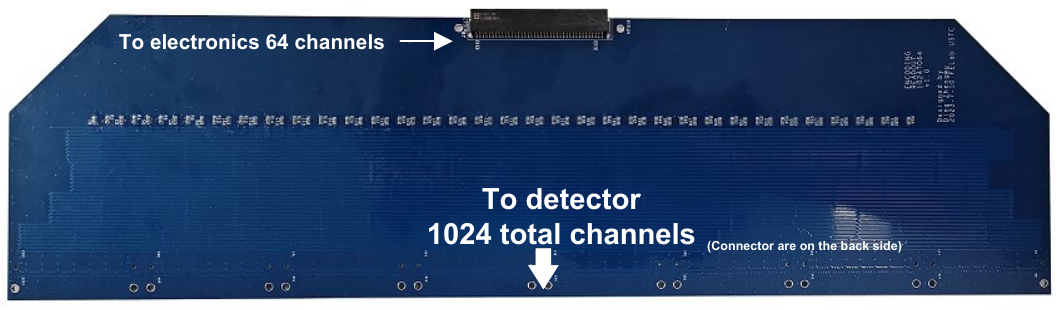}}
\caption{Circuit multiplexing 1024 detector strips to 64 readout electronics using the complete graph method.}
\label{fig5}
\end{figure}

Three types of encoded multiplexing circuits were designed based on the mathematical model described earlier. 
Two of these circuits multiplex 512 detector channels into 64 readout electronics channels, utilizing Hamiltonian paths of complete graph and bipartite graph, respectively. 
A photograph of these circuits is shown in \figurename~\ref{fig4}.
The third circuit multiplexes 1024 detector channels into 64 readout electronics channels using Hamiltonian paths of complete graph, as shown in \figurename~\ref{fig5}.

\subsection{Experimental Setup}
The front-end electronics card (FEC) for reading out detector signals utilizes an ASIC called ASTRE (ASIC with SCA \& Trigger for Detector Readout Electronics)~\cite{bib_astre_chip}. 
The ASTRE chip, developed by CEA Saclay, is an upgraded version of the AGET (ASIC for General Electronics for TPC) chip~\cite{bib:bib_AGET}.
It is an analog readout chip with 64 channels. 
Each channel consists of a charge-sensitive amplifier (CSA), a pole-zero cancellation (PZC) circuit, a Sallen-Key (SK) filter, a 512-cell switched capacitor array, and a discriminator for trigger generation.

The FEC design utilized in the test is derived from the PandaX-III (Particle AND Astrophysical Xenon experiment III) project and is compatible with the ASTRE chips.
Each FEC integrates four ASTRE chips, offering a total of 256 input channels.
The input part of the FEC includes electrostatic discharge (ESD) protection circuits and AC coupling capacitors.  
Analog signals are amplified and sampled by the ASTRE chips, and their outputs are digitized by single-channel 12-bit ADCs (analog-to-digital converters).
Upon receiving a valid trigger signal, the FPGA on the FEC halts the acquisition phase of the ASTRE chips after a programmable delay and initiates ADC conversion. 
Further details about the FEC can be found in~\cite{bib:bib_FEC}, where the ASIC used has been upgraded to ASTRE chip.
A DAQ board equipped with multiple fiber-optic interfaces is utilized for data acquisition. It performs functions such as data collection, trigger and clock distribution, trigger judgment, and pre-processing of acquisition data. Further details about the DAQ board can be found in~\cite{bib_daq}.

\subsection{Impact of the Multiplexing Circuit on FEC Noise and Gain}
The noise of the CSA is influenced by the input capacitance, which is affected by the trace length of the multiplexing circuit and the merging of strips in the Micromegas detectors.

The root mean square (RMS) of the equivalent noise charge (ENC) of the FEC was measured under the same conditions using different multiplexing circuits to connect the FEC to the detector.
To compare this with the direct readout method, in which each strip of the detector is read out by a single electronics channel, a direct readout connector was designed with a trace length comparable to that of the multiplexing circuit.
During the tests, the feedback capacitance of the CSA in the ASIC was set to $120~\mathrm{fF}$, corresponding to a dynamic range of 0 to 120 fC, and the shaping time was set to $1039~\mathrm{ns}$.

\begin{figure}[htbp]
\centerline{\includegraphics[width=3.5in]{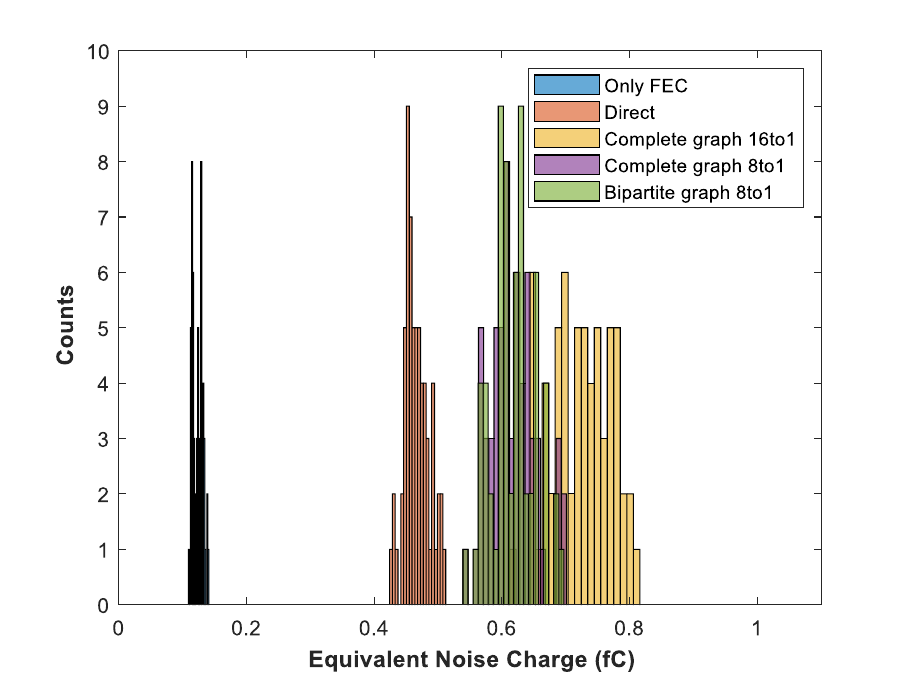}}
\caption{RMS of ENC measurements for different multiplexing circuits, compared to the direct readout method.}
\label{fig6}
\end{figure}

As shown in~\figurename~\ref{fig6}, when the detector was connected using the direct readout board, the RMS of the ENC increased from 0.12 fC to 0.45 fC compared to the bare FEC test. 
Similarly, when the multiplexing circuits were applied, the noise level increased, with the maximum RMS of the ENC measured at 0.81 fC under a compression ratio of 16. 
Nevertheless, the incremental noise remains within an acceptable range.

The influence of the input capacitor can be attributed to the capacitance introduced by the PCB traces and the capacitance resulting from the merging of multiple detector strips. 
To isolate these two factors, a test was conducted in which the ENC and input capacitance were measured by connecting different numbers of connectors to the detector.
During the test, the capacitance from the PCB traces was kept constant, while the effect of merging the strips was varied.
The input capacitance was measured using a Uni-Trend UT116 Series SMD tester. 
\figurename~\ref{fig9}(a) shows the input capacitance for different compression ratios. 
Here, a compression ratio of zero indicates that the circuit is not connected to the detector, and the measured value corresponds only to the capacitance of the traces on the multiplexing board. 
From the results, the detector capacitance is approximately $100~\mathrm{pF/channel}$, which aligns well with the values observed in the three circuits tested.
The RMS of ENC for different multiplexing ratios is shown in \figurename~\ref{fig9}(b). With increasing compression ratios, the ENC increased gradually, indicating that the influence of the input capacitance was acceptable.

The gain calibration of the FEC was also tested with different multiplexing circuits. 
The amplitude of the CSA can be expressed as:
\begin{equation}
\begin{aligned}
\dot{\mathrm{V_o}} \approx \frac{\dot{A}_{v o}Q}{C_{D}+(1+\dot{A}_{v o}){C_{F}}}  
\end{aligned}
\label{eq_csa}
\end{equation}
where $Q$ is the charge generated by the detector, $\dot{A}_{v o}$ is the open-loop gain of the amplifier used in the CSA, $C_F$ is the feedback capacitor value of the CSA, and $C_D$ is the total input capacitance seen by the CSA. 
\begin{figure}[htbp]
\centerline{\includegraphics[width=3.5in]{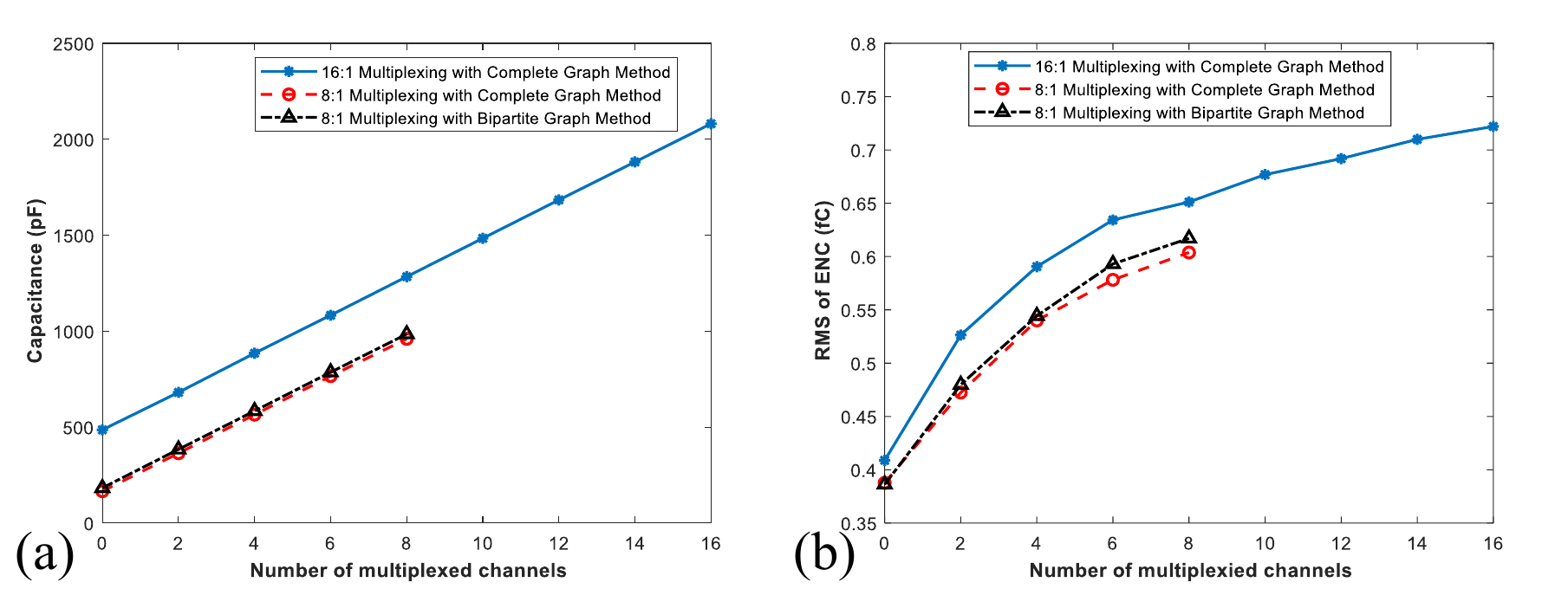}}
\caption{Test results of different compression ratios (zero means not connected to the detector). (a) Input capacitance for different compression ratios. (b) ENC with different compression ratios.}
\label{fig7}
\end{figure}

\begin{figure}[htbp]
\centerline{\includegraphics[width=3.5in]{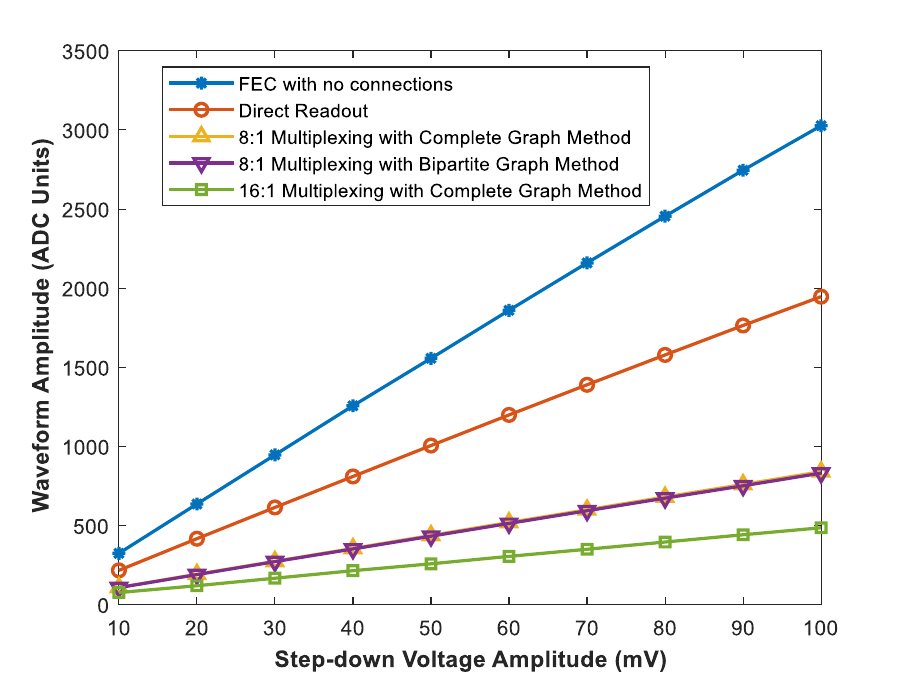}}
\caption{Calibration results of the FEC under different multiplexing circuits. A 1 pF capacitor (5\% accuracy) was used to inject a charge signal, and the amplitude of the recorded waveform was used to calculate the gain.}
\label{fig8}
\end{figure}

A 1 pF capacitor (5\% accuracy) was used to inject a charge signal into the input channel of the FEC by applying a step-down voltage to the capacitor, and the amplitude of the waveform recorded by the FEC was calculated.
The calibration curve of the FEC under different multiplexing circuits is shown in \figurename\ref{fig8}
The gain loss is attributed to the large capacitance of the strips and traces.
Although the signal-to-noise (S/N) ratio is reduced, this can be compensated by increasing the gain of the Micromegas detectors, allowing them to maintain relatively high detection efficiency.

\subsection{Detection Efficiency and Spatial Resolution of Micromegas}
A cosmic-ray muon test platform was constructed, consisting of six layers of reference detectors and one layer of target detector. 
The six reference detector layers were divided into two groups: three layers positioned above the target detector and three layers below it. 
The working gas is mixture of argon and CO2 (7\%) and the working voltages of the mesh and drift cathode were $\mathrm{-570 ~V}$ and $\mathrm{-720 ~V}$, respectively.

\begin{figure}[htbp]
\centerline{\includegraphics[width=3.5in]{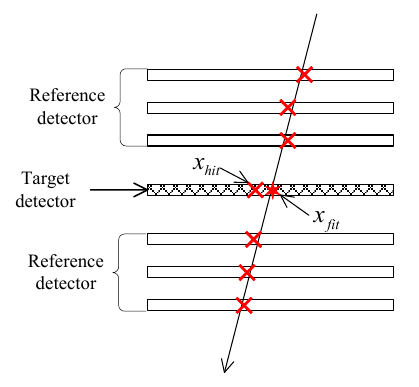}}
\caption{Illustration of the method for calculating spatial resolution and detection efficiency. The actual hit positions are represented by red crosses, while the fitted hit position is indicated by a red hexagram.}
\label{fig9}
\end{figure}

The system worked in a self-trigger mode, where the six layers of reference detectors generate over-threshold signals within a $1 ~ \mu \mathrm{s}$ time window, and the DAQ board judges whether a valid event is detected. 
The target detector, connected to the FEC with different types of multiplexing circuits and a direct readout circuit, was not used for trigger generation.
The method for calculating spatial resolution and detection efficiency is illustrated in \figurename~\ref{fig9}. 
Using the hit positions of the reference detectors, a reference tracker was fitted.
The inferred hit position, $x_{fit}$, was calculated based on the measured and aligned Z positions of the detectors.

\begin{figure}[htbp]
\centerline{\includegraphics[width=3.5in]{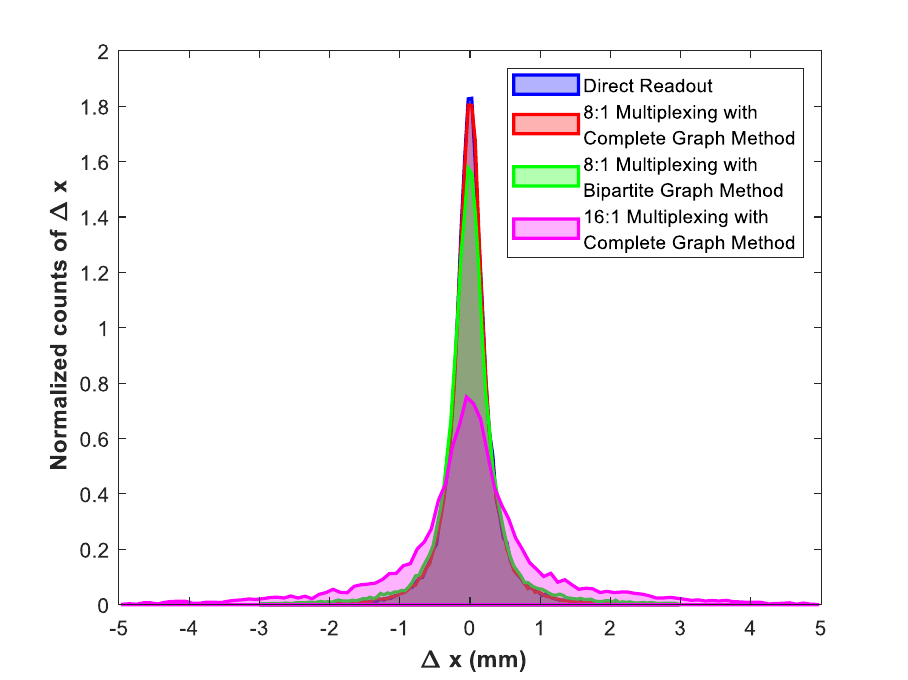}}
\caption{Normalized distribution of the deviation $\Delta x=x_{hit} - x_{fit}$ for the target detector.}
\label{fig10}
\end{figure}

For spatial resolution calculation, the deviation $\Delta x=x_{hit} - x_{fit}$ was calculated, where $x_{hit}$ is the hit position measured by the target detector.
An alignment procedure was applied to correct for installation offsets and rotations.
By statistically analyzing a large number of cosmic-ray muon tracks, the distribution of $\Delta x$ was obtained, and its standard deviation was taken as the spatial resolution of the target detector. 
For each event, the charge signals were distributed across several strips, and a charge centroid method was used to reconstruct the hit position. 
The normalized distribution of $\Delta x$ is shown in \figurename\ref{fig10}.
For circuits multiplexing 512 detector strips to 64 readout electronics channels using the complete graph and bipartite graph methods, the spatial resolution results are comparable to that of the direct readout method. 
However, for the circuit multiplexing 1024 detector strips to 64 readout electronics channels, the spatial resolution degrades. 
This degradation is attributed to charge loss on some strips, which increases the uncertainty in position reconstruction.

For detection efficiency calculation, two counters are used: one to count the total number of reference trackers ($N_{total}$) and the other to count the successfully reconstructed hit positions in the target detector ($N_{target}$). 
For each reference tracker, $N_{total}$ is incremented by 1.
If the target detector reconstructed a hit position with $\Delta x$ less than $10\cdot \sigma$ where $\sigma$ is the spatial resolution of the target detector, $N_{target}$ is incremented by 1.
The detection efficiency is then calculated as $N_{target}/N_{total}$.
The detection efficiency for the direct readout circuit is 96.7\%.
For the circuits multiplexing 512 detector strips to 64 readout electronics channels using the complete graph and bipartite graph methods, the detection efficiencies are 94.5\% and 94.3\%, respectively. 
For the circuit multiplexing 1024 detector strips to 64 readout electronics channels, the detection efficiency is 93.6\%.
Although the detection efficiency decreases due to charge loss caused by the increased input capacitance of the multiplexing circuit, the high compression ratio significantly reduces the required number of readout electronics. 
In experiments involving large-area detectors, this trade-off is acceptable.

\subsection{Applied in Cosmic-Ray muon Image}
The multiplexing circuits, which multiplex 512 detector strips into 64 readout electronics channels using the complete graph method, were implemented in a muon imaging facility.
This facility consists of four layers of Micromegas detectors along with their corresponding FECs. 
In total, 8000 strips are read out through 1024 electronics channels via the multiplexing circuits. 
The detectors were operated under the same conditions (voltage and working gas) as described in earlier performance tests.
The facility was configured to operate in a self-trigger mode, where a valid trigger signal was generated when at least three out of the four detectors produced signals exceeding the predefined threshold.

\begin{figure}[htbp]
\centerline{\includegraphics[width=3.5in]{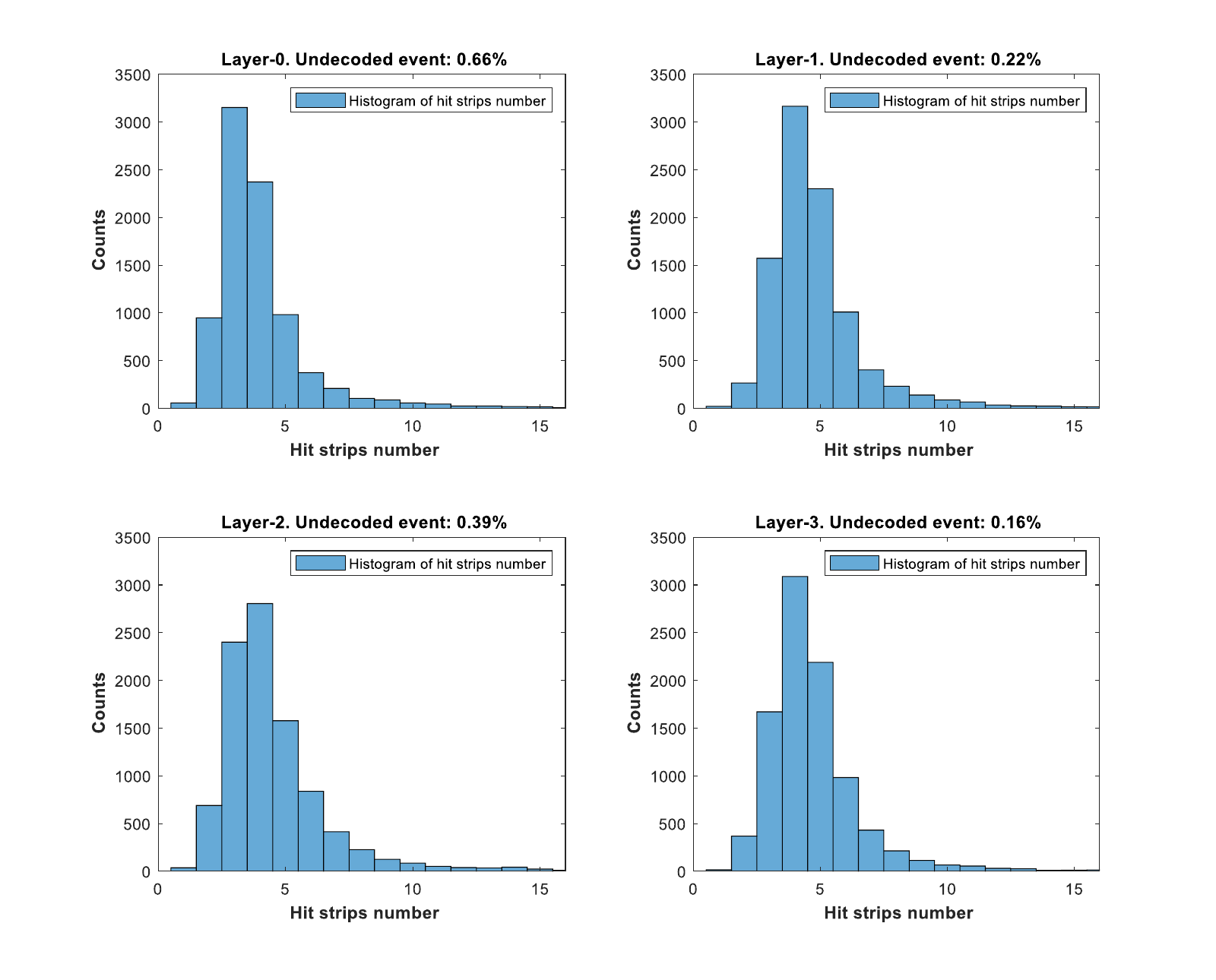}}
\caption{Histogram of the number of hit channels per event, showing that single-channel events account for less than 1\%.}
\label{fig11}
\end{figure}

The histogram of the number of hit strips per event is shown in \figurename~\ref{fig11}. 
For successful hit position reconstruction via multiplexing, at least two channels must record signals. 
Single-channel hit events cannot be decoded to their original hit positions, and in our test, these events accounted for less than 1\% of the total. 
This result indicates that for a Micromegas detector with a $400~\mathrm{\mu m}$ strip pitch and a $5~\mathrm{mm}$ drift region thickness, the diffusion of primary ionization typically spreads across multiple strips. Therefore, the design requirements of the multiplexing circuits are satisfied.

\begin{figure}[htbp]
\centerline{\includegraphics[width=3.5in]{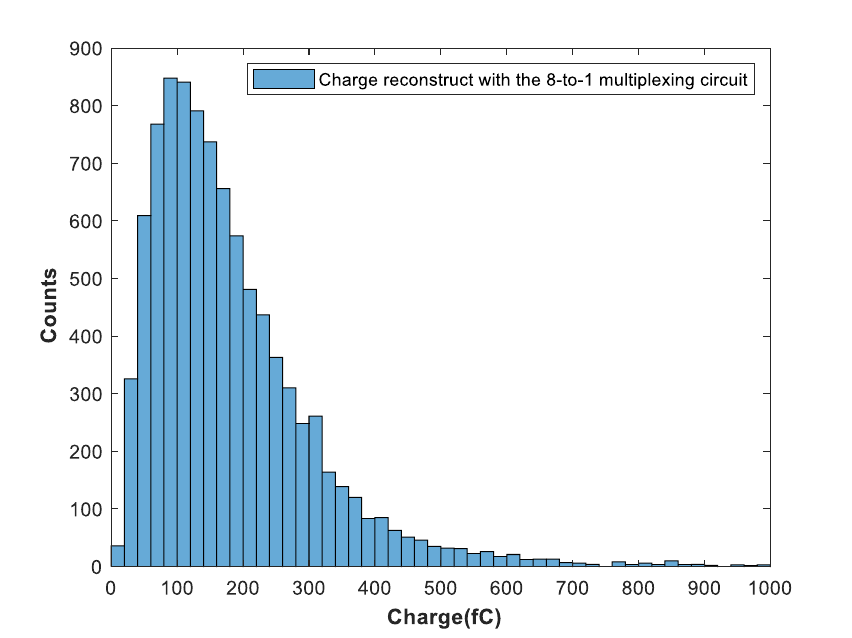}}
\caption{Reconstructed charge measured by the FEC, following a convolution of Landau and Gaussian distributions.}
\label{fig12}
\end{figure}

Regarding charge reconstruction, \figurename~\ref{fig12} shows the reconstructed charge measured by the FEC. 
This charge corresponds to the energy deposited by cosmic-ray muons in the Micromegas detector and follows a convolution of Landau and Gaussian distributions.

\section{Conclusion and Discussion}
This paper presents two novel multiplexing methods based on the developed Eulerian circuit theory. 
These multiplexing techniques are intended for application in micro-pattern gaseous detectors to reduce the number of electronics channels.
By employing these methods, we proposed a general framework for generating multiplexing schemes, and three types of multiplexing circuits were successfully designed and implemented. 
The highest compression ratio achieved is 16:1, enabling the readout of 1024 detector channels with only 64 electronics channels.
In combination with front-end electronics and Micromegas detectors, the performance of these circuits was verified, demonstrating their effectiveness in reliably reading detector signals with high accuracy and efficiency.

Despite these promising results, the gain of the charge-sensitive amplifier is reduced due to the increased input capacitance, which in turn reduces the signal-to-noise ratio.
Currently, this issue is mitigated by increasing the avalanche gain of the Micromegas detectors. 
Future research should prioritize enhancing the open-loop gain of the CSA amplifier, as outlined in (\ref{eq_csa}), to further improve performance.

In conclusion, this study presents a novel method for reading out detectors with fine readout units, achieving a significant reduction in the number of electronics channels while maintaining fine spatial resolution and high detection efficiency.
The proposed approach not only demonstrates its potential application in cosmic-ray muon imaging but also provides a flexible framework for future advancements in high-density detector readout systems.


\begin{thebibliography}{00}
\bibitem{bib_gem} F. Sauli, “GEM: A new concept for electron amplification in gas detectors,” Nuclear Instruments and Methods in Physics Research Section A: Accelerators, Spectrometers, Detectors and Associated Equipment, vol. 386, no. 2, pp. 531–534, Feb. 1997, doi: 10.1016/S0168-9002(96)01172-2.

\bibitem{bib_mm} Y. Giomataris, Ph. Rebourgeard, J. P. Robert, and G. Charpak, “MICROMEGAS: a high-granularity position-sensitive gaseous detector for high particle-flux environments,” Nuclear Instruments and Methods in Physics Research Section A: Accelerators, Spectrometers, Detectors and Associated Equipment, vol. 376, no. 1, pp. 29–35, Jun. 1996, doi: 10.1016/0168-9002(96)00175-1.

\bibitem{bib_thgem} A. Breskin, R. Alon, M. Cortesi, R. Chechik, J. Miyamoto, V. Dangendorf, et al., ``A concise review on THGEM detectors,' Nuclear Instruments and Methods in Physics Research Section A: Accelerators, Spectrometers, Detectors and Associated Equipment, vol. 598, no. 1, pp. 107–111, Jan. 2009, doi: 10.1016/j.nima.2008.08.062.

\bibitem{bib_rwell} G. Bencivenni, R. D. Oliveira, G. Morello, and M. P. Lener, “The micro-Resistive WELL detector: a compact spark-protected single amplification-stage MPGD,” J. Inst., vol. 10, no. 02, pp. P02008–P02008, Feb. 2015, doi: 10.1088/1748-0221/10/02/P02008.

\bibitem{Perez_Lara_a_comparative_study_of_straight_strip_and_zigzag_interleaved} C. Perez-Lara, S. Aune, B. Azmoun, K. Dehmelt, A. Deshpande, W. Fan, et al., “A Comparative Study of Straight-Strip and Zigzag-Interleaved Anode Patterns for MPGD Readouts,” IEEE Transactions on Nuclear Science, vol. 69, no. 1, pp. 50–55, Jan. 2022, doi: 10.1109/TNS.2021.3132946.

\bibitem{bib_procureur_genatic_multiplexing} Procureur S., Dupré R., and Aune S., “Genetic multiplexing and first results with a 50×50cm2 Micromegas,” Nuclear Instruments and Methods in Physics Research Section A: Accelerators, Spectrometers, Detectors and Associated Equipment, vol. 729, pp. 888–894, Nov. 2013, doi: 10.1016/j.nima.2013.08.071.

\bibitem{bib_Yue_an_encoding_readout_method} X. Yue, M. Zeng, Y. Wang, X. Wang, Z. Zeng, Z. Zhao, et al., “An encoding readout method used for Multi-gap Resistive Plate Chambers (MRPCs) for muon tomography,” J. Inst., vol. 9, no. 09, pp. C09033–C09033, Sep. 2014, doi: 10.1088/1748-0221/9/09/C09033.

\bibitem{bib_Yue_Mathematical_modelling_and_study} X. Yue, M. Zeng, Z. Zeng, Y. Wang, X. Wang, Z. Zhao, et al., “Mathematical modelling and study of the encoding readout scheme for position sensitive detectors,” Nuclear Instruments and Methods in Physics Research Section A: Accelerators, Spectrometers, Detectors and Associated Equipment, vol. 816, pp. 33–39, Apr. 2016, doi: 10.1016/j.nima.2016.01.080.

\bibitem{bib_BQi_A_novel_method_of_encoded_multiplexing_readout_2016} B. Qi, S. Liu, H. Ji, Z. Shen, S. Ma, H. Liu, et al., “A novel method of encoded multiplexing readout for micro-pattern gas detectors,” Chinese Phys. C, vol. 40, no. 5, p. 056102, May 2016, doi: 10.1088/1674-1137/40/5/056102.

\bibitem{bib_GYuan_2D_encoded_multiplexing_readout_for_thgem} G. Yuan, S. Liu, B. Qi, C. Feng, and S. Ma, “2-D Encoded Multiplexing Readout for THGEM,” IEEE Trans. Nucl. Sci., vol. 64, no. 6, pp. 1346–1349, Jun. 2017, doi: 10.1109/TNS.2017.2706766.

\bibitem{bib_JPan_position_encoding_readout_electronics_of_large_area2019} J. Pan, C. Feng, Z. Zhang, Y. Wang, H. Chen, J. Wang, et al., “Position Encoding Readout Electronics of Large Area Micromegas Detectors aiming for Cosmic Ray Muon Imaging,” in 2019 IEEE Nuclear Science Symposium and Medical Imaging Conference (NSS/MIC), Oct. 2019, pp. 1–5. doi: 10.1109/NSS/MIC42101.2019.9060024.

\bibitem{bib_JLiu_An_encoding_readout_scheme_for_micromegas_detector_used_in_muography} J. Liu, S. Liu, Y. Liu, Y. Wang, Z. Shen, C. Feng, et al., “An Encoding Readout Scheme for Micromegas Detector Used in Muography,” in 2021 IEEE Nuclear Science Symposium and Medical Imaging Conference (NSS/MIC), Oct. 2021, pp. 1–4. doi: 10.1109/NSS/MIC44867.2021.9875710.

\bibitem{bib_G_aielli_The_rpc_space_resolution_2014} G. Aielli, R. Cardarelli, L. Di Stante, B. Liberti, L. Paolozzi, E. Pastori, et al., “The RPC space resolution with the charge centroid method,” J. Inst., vol. 9, no. 09, pp. C09030–C09030, Sep. 2014, doi: 10.1088/1748-0221/9/09/C09030.

\bibitem{bib_MIodice_Performance_studies_ofMicromegas_2014} M Iodice, “Performance studies of MicroMegas for the ATLAS experiment,” J. Inst., vol. 9, no. 01, pp. C01017–C01017, Jan. 2014, doi: 10.1088/1748-0221/9/01/C01017.

\bibitem{bib_LLavezzi_Standalone_codes_for_simulation_and_reconstruction_2020} L. Lavezzi, M. Alexeev, A. Amoroso, S. Bagnasco, R. Baldini Ferroli, I. Balossino, et al., “Standalone codes for simulation and reconstruction of a triple-GEM: GTS and GRAAL,” J. Phys.: Conf. Ser., vol. 1561, no. 1, p. 012014, Jun. 2020, doi: 10.1088/1742-6596/1561/1/012014.

\bibitem{bib_DRBean_Recursive_Euler_and_hamilton_paths_1976} D. R. Bean, “Recursive Euler and Hamilton paths,” Proc. Amer. Math. Soc., vol. 55, no. 2, pp. 385–394, 1976, doi: 10.1090/S0002-9939-1976-0416888-0.


\bibitem{bib_Reinhard_Diestel_Graph_Theory_2017} Reinhard Diestel, Graph Theory, 5th ed. in Graduate Texts in Mathematics. Berlin Heidelberg: Springer-Verlag, 2017. doi: 10.1007/978-3-662-53622-3.

\bibitem{bib_Jimbo_Shuji_On_the_Eulerian_recurrent_lengths_of_complete_bipartite_graphs_and_complete_graphs_2014} Jimbo Shuji, “On the Eulerian recurrent lengths of complete bipartite graphs and complete graphs,” IOP Conf. Ser.: Mater. Sci. Eng., vol. 58, p. 012019, Jun. 2014, doi: 10.1088/1757-899X/58/1/012019.

\bibitem{bib_Jimbo_Shuji_The_eulerian_recurrent_length_of_a_complete_graphs_2012} Jimbo Shuji, “The Eulerian Recurrent Lengths of Complete Graphs,” 2012. [Online]. Available: https://www.kurims.kyoto-u.ac.jp/~kyodo/kokyuroku/contents/pdf/1809-02.pdf. Accessed: May 16, 2024.


\bibitem{bib_LMetcalf_Chapter5_graph_theory_2016} L. Metcalf and W. Casey, “Chapter 5 - Graph theory,” in Cybersecurity and Applied Mathematics, L. Metcalf and W. Casey, Eds., Boston: Syngress, 2016, pp. 67–94. [Online]. Available: https://www.sciencedirect.com/science/article/pii/B9780128044520000051. Accessed: Oct. 10, 2024.

\bibitem{bib_MyGithub} “wyu0725/encoded\_multiplexing\_scheme: This project is about the construction of the encoded multiplexing method.” [Online]. Available: \href{https://github.com/wyu0725/encoded_multiplexing_scheme}{https://github.com/wyu0725/encoded\_multiplexing\_scheme}. Accessed: May 16, 2024.

\bibitem{bib_Jianxingfeng_A_themal_bonding_method_for_manufacturing_Micromegas_detector_2021} J Feng, Z Zhang, J Liu, B Qi, A Wang, M Shao, et al., “A thermal bonding method for manufacturing Micromegas detectors,” Nuclear Instruments and Methods in Physics Research Section A: Accelerators, Spectrometers, Detectors and Associated Equipment, vol. 989, p. 164958, Feb. 2021, doi: 10.1016/j.nima.2020.164958.

\bibitem{bib_astre_chip} D. Baudin, D. Attié, P. Baron, D. Bernard, D. Calvet, A. Delbart, et al., “ASTRE: ASIC with switched capacitor array (SCA) and trigger for detector readout electronics hardened against Single Event Latchup (SEL),” Nuclear Instruments and Methods in Physics Research Section A: Accelerators, Spectrometers, Detectors and Associated Equipment, vol. 912, pp. 66–69, Dec. 2018, doi: 10.1016/j.nima.2017.10.043.

\bibitem{bib:bib_AGET} S. Anvar, P. Baron, B. Blank, J. Chavas, E. Delagnes, F. Druillole, et al., “AGET, the GET front-end ASIC, for the readout of the Time Projection Chambers used in nuclear physic experiments,” in 2011 IEEE Nuclear Science Symposium Conference Record, Oct. 2011, pp. 745–749. doi: 10.1109/NSSMIC.2011.6154095.

\bibitem{bib:bib_FEC} D. Zhu, S. Liu, C. Feng, C. Li, J. Dong, H. Chen, et al., “Development of the Front-End Electronics for PandaX-III Prototype TPC,” IEEE Transactions on Nuclear Science, vol. 66, no. 7, pp. 1123–1129, Jul. 2019, doi: 10.1109/TNS.2019.2907125.


\bibitem{bib_daq} J. Liu, Y. Wang, C. Feng, S. Liu, and Q. Chen, “A Back-End Electronics Based on Fiber Communication for Small to Medium-Scale Physics Experiments,” Jul. 09, 2024, arXiv: arXiv:2407.06786. doi: 10.48550/arXiv.2407.06786.


\end{thebibliography}
\end{document}